\documentclass[aps,pre,twocolumn,groupedaddress,superscriptaddress,showpacs]{revtex4-1}

\usepackage{amsmath}
\usepackage{graphicx}
\usepackage{float}
\usepackage{color}

\usepackage[utf8]{inputenc}
\makeatletter
\def\qrr@split@result#1 #2\@qrr@split@result{\edef\erfInput{#1}\edef\erfResult{#2}}
\newcommand*{\gnuplotErf}[2][\jobname.eval]{%
    \immediate\write18{gnuplot -e "set print '#1'; print #2, erf(#2);"}%
    \everyeof{\noexpand}
    \edef\qrr@temp{\input{#1} }%
    \expandafter\qrr@split@result\qrr@temp\@qrr@split@result
}
\makeatother
\DeclareMathOperator{\erf}{erf}

\begin{document}
	
	\title{Dynamical demixing of a binary mixture under sedimentation} 
	
	\author{Andr\'e S. Nunes}
	\affiliation{Centro de F\'isica Te\'orica e Computacional, Faculdade de Ciências, Universidade de Lisboa,
		P-1749-016 Lisboa, Portugal}
    \affiliation{Departamento de F\'{\i}sica, Faculdade
		de Ci\^{e}ncias, Universidade de Lisboa, P-1749-016 Lisboa, Portugal}
	\affiliation{Electric Ant Lab, Science Park 106, 1098 XG, Amsterdam, The Netherlands}
	
    \author{Rodrigo C. V. Coelho}
	\affiliation{Centro de F\'isica Te\'orica e Computacional, Faculdade de Ciências, Universidade de Lisboa,
		P-1749-016 Lisboa, Portugal}
    \affiliation{Departamento de F\'{\i}sica, Faculdade
		de Ci\^{e}ncias, Universidade de Lisboa, P-1749-016 Lisboa, Portugal}	
	
	\author{Vasco C. Braz}
	\affiliation{Centro de F\'isica Te\'orica e Computacional, Faculdade de Ciências, Universidade de Lisboa,
		P-1749-016 Lisboa, Portugal}
    \affiliation{Departamento de F\'{\i}sica, Faculdade
		de Ci\^{e}ncias, Universidade de Lisboa, P-1749-016 Lisboa, Portugal}
		
	\author{Margarida M. Telo da Gama}
	\affiliation{Centro de F\'isica Te\'orica e Computacional, Faculdade de Ciências, Universidade de Lisboa,
		P-1749-016 Lisboa, Portugal}
    \affiliation{Departamento de F\'{\i}sica, Faculdade
		de Ci\^{e}ncias, Universidade de Lisboa, P-1749-016 Lisboa, Portugal}
	
	\author{Nuno A. M. Ara\'ujo}
	\email{nmaraujo@fc.ul.pt}
	\affiliation{Centro de F\'isica Te\'orica e Computacional, Faculdade de Ciências, Universidade de Lisboa,
		P-1749-016 Lisboa, Portugal}
    \affiliation{Departamento de F\'{\i}sica, Faculdade
		de Ci\^{e}ncias, Universidade de Lisboa, P-1749-016 Lisboa, Portugal}
	
	\begin{abstract}
		We investigate the sedimentation dynamics of a binary mixture, the species of which differ by their Stokes coefficients but are identical otherwise. We analyze the sedimentation dynamics and the morphology of the final deposits using Brownian dynamics simulations for mixtures with a range of sedimentation velocities of both species. In addition, we use the lattice Boltzmann method to study hydrodynamic effects. We found a threshold in the difference of the sedimentation velocities above which the species in the final deposit are segregated. The degree of segregation increases with the difference in the Stokes coefficients or the sedimentation velocities above the threshold. We propose a simple analytical model that captures the main features of the simulated deposits. 
	\end{abstract}
	
	\maketitle
	
	\section{Introduction}

	The process of sedimentation, where particles in suspension settle in the presence of a gravitational field, is ubiquitous over a wide range of length scales \cite{Piazza2014, Cho2011, Park2012, Yoder2001, Schuck2003}. For example, sedimentation plays a relevant role in natural water transport, affecting the chemical composition of the seabed \cite{Smith1986} and the water quality in reservoirs \cite{Knox2006, Holz2004}. At the other end of the scale, sedimentation by ultracentrifugation is used as an analytical tool in medical, biological and pharmaceutical applications, where the constituents of a suspension are separated by molecular weight \cite{Cole2008, Lebowitz2002}. At the fundamental level, sedimentation experiments were developed and used extensively in statistical physics and colloidal science to evaluate the equation of state of hard spheres \cite{Piazza1993} and to study the phase diagram of colloidal particles \cite{Buzzaccaro2008}. 
	
	Studies of the sedimentation of mixtures of particles that differ in their buoyant mass revealed a rich phase stacking diagram under thermodynamic equilibrium conditions \cite{Batchelor1986, Heras2013}. The structure of the final deposit depends not only on the difference in buoyant masses but also on the particle-particle interactions \cite{Araujo2017,Rasa2004, Heras2012, Kleshchanok2012,Dias2014}. The roughness of the particle surface is known to affect the hydrodynamics of the surrounding fluid, e.g., alters the lubrication film thickness \cite{Wilson2002,Ilhan2019}. These conditions could be realized in an experiment with particles composed by a rigid core covered by an elastic surface layer, which would affect the hydrodynamics while the pairwise interactions remain dominated by the rigid cores. In order to shed light on the role of the hydrodynamic radius on the sedimentation dynamics, we consider a binary mixture of particles that differ only through their Stokes coefficient when using molecular dynamics. Following a method developed previously \cite{Nunes2018}, we consider that the particles differ by their Stokes coefficients only, and are identical otherwise. Thus, the thermodynamic phase is perfectly mixed and demixing, if it occurs, is dynamically driven. Hydrodynamic interactions are know to be relevant in different limits, triggering, for example, a number of different instabilities~\cite{Nitsche1997,Lowen2010}. To account for hydrodynamic effects, we consider a complementary set of simulations by modelling the fluid-particle interaction with the lattice Boltzmann method (LBM). 
	
	When thermal fluctuations are negligible, colloidal particles in solution are expected to sediment with a sedimentation velocity that depends only on the strength of the gravitational field and their Stokes coefficient. Thus, distinct Stokes coefficients imply different sedimentation velocities. In what follows, we show that the morphology of the final deposit depends crucially on the ratio of the sedimentation velocities. Above a certain threshold, which will be quantified below, the particles are segregated in the final deposit, as they arrive at the substrate at different rates and do not have time to relax to the thermodynamic equilibrium mixed state. We investigate this segregation and discuss its dependence on the model parameters.
	
	The paper is organized in the following way. In Section \ref{sec::model}, we describe the model and the details of the simulations. Results from the particle-based simulations (Brownian dynamics), LBM and an analytical model are discussed in Section \ref{sec::results}. Finally, we draw some conclusions in Section \ref{sec::conclusions}. 
	
	\section{Model and Simulations~\label{sec::model}}

	\subsection{Molecular dynamics}
	We consider a binary mixture of identical spherical particles where the two species are characterized by distinct Stokes coefficients. The particles are in a uniform gravitational field along the vertical direction ($y$-direction) and inside a rectangular two-dimensional box of width $L_x$ and height $L_y$. The boundary conditions are periodic in the $x$-direction and are rigid walls in the $y$-direction. The trajectory of each particle $i$ is obtained by solving the Langevin equation, in the overdamped regime,
	\begin{equation}
	\gamma_i \frac{d\vec{r}_i}{dt} = -\nabla _i\left[\sum_{j}^{N_t} V_{ij}(r) \right] +m\vec{g}+ \vec{\xi}_{i},\quad j\neq i\label{Eq:MotionEq}
	\end{equation}
	where, $\vec{r}_i$ is the position of particle  $i$, $V_{ij}$ is the pairwise potential, $N_t$ the total number of particles, $m$ is a parameter that takes into account the effects of mass and buoyancy of the particles, $\vec{g}=-g\vec{e}_y$ the gravitational field, $\vec{\xi}_i$ a stochastic force, and $\gamma_i$ is the Stokes coefficient. The two species differ through the values of $\gamma_i$: $\gamma_f$ for fast particles and $\gamma_s$ for slow ones, such that $\gamma_f<\gamma_s$. The diffusion coefficient of each species is also different, as given by the Stokes-Einstein relation $D_i=k_BT/\gamma_i$. As a result, the two species have different sedimentation velocities, $\vec{v}_{i}=\frac{m}{\gamma_i} \vec{g}$ since the the particles have the same mass. The fluid is in thermodynamic equilibrium at a thermostat temperature $T$ and hydrodynamic effects are neglected. Thus the time series of the stochastic force is drawn from a Gaussian distribution with zero mean and uncorrelated second moments in time and space, given by $\left\langle \xi_i^k(t)\xi_i^l(t')\right\rangle = 2k_BT\gamma_i\delta_{kl}\delta(t-t')$, where $k$ and $l$ refer to the coordinates of the vector $\vec{\xi}_i$. 
	
	In order to focus on purely dynamical effects, we consider that the particle-particle interactions are identical for all the particles. We describe this interaction through a (repulsive) Lennard-Jones potential, truncated at a cut-off distance $r_{cut}=2^{\frac{1}{6}}\sigma$,
	
	\begin{equation}
	V_{ij}(r)=\epsilon\left[ \left( \frac{\sigma}{r}\right) ^{12}-\left( \frac{\sigma}{r}\right) ^{6}\right], \label{Lennard-jones}
	\end{equation}
	where $\epsilon$ sets the energy scale and $\sigma$ the size of the particles. Thus, the potential depends only on the distance $r=\vert\vec{r}_i-\vec{r}_j\vert$ between the particles $i$ and $j$.
	
	Hereafter, $\sigma$ sets the unit of length. The energy is expressed in units of $k_BT$, time is defined in units of the Brownian time $\tau=\sigma^2\gamma (k_BT)^{-1}$ and the strength of the external field, $g$, is given in units of $k_BT/(m \sigma)$. Equation \ref{Eq:MotionEq} is integrated using a second-order stochastic Runge-Kutta numerical scheme, proposed by Bra\'nka and Heyes ~\cite{Branka1999}, with a time-step of $\Delta t=10^{-4}\tau$. Initially, the particles are distributed uniformly at random (without overlapping) in the simulation box. Unless stated otherwise, we set $\epsilon=1$ and $g=12$. The box size is $L_y=200$ and $L_x=37.5$ and the binary mixture consists of $N=3000$ particles, with $N/2$ particles of each species. The initial number density is $\rho_0 = 0.4$.

	\subsection{Lattice Boltzmann method}
	\label{lbm-intro-sec}
	
To account for hydrodynamic effects, we employ a Lattice Boltzmann method~\cite{Succi2018,Krger2017} to describe the motion of the suspended medium and couple it to a discrete element simulation of the particles. For the fluid, the position $\vec{x}$ is discretized on a regular lattice with lattice spacing $\Delta x$ and the velocity of the particles in the suspending fluid is discretized according to the D3Q19 lattice (with 19 vectors in three dimensions): the weights for the different velocity vectors are $w_\alpha=1/3$ for $\vert \mathbf{c} \vert^2=0$, $w_\alpha=1/18$ for $\vert \mathbf{c} \vert^2=1\Delta x$ and $w_\alpha=1/36$ $\vert \mathbf{c} \vert^2=2\Delta x$, where $\alpha=1, \ldots, 19$ labels the velocity vector. The speed of sound in this lattice is $c_s=1/\sqrt{3}$. The LBM is based on solving the dynamics of the distribution function $f_\alpha$, which is governed by the Boltzmann equation. Its discretized version with the single relaxation time collision operator is:
\begin{equation}
 f_\alpha (\vec{x}+\vec{c}_\alpha, t+\Delta t)  -  f_\alpha(\vec{x},  t) =  - \frac{f_\alpha - f^{eq}_\alpha}{\tau_f}\Delta t ,
 \label{boltz-force-eq}
\end{equation}
where $\Delta t$ is the time step and $\tau_f$ is the relaxation time which sets the kinematic viscosity of the fluid $\nu=c_s^2(\tau_f-1/2)$. The equilibrium distribution function is given by:
\begin{equation}
 f^{eq}_\alpha = \rho w_\alpha \left[ 1 + \frac{\vec{c}_\alpha \cdot \vec{u}}{c_s^2} - \frac{u^2}{2c_s^2} + \frac{(\vec{c}_\alpha \cdot \vec{u})^2}{2c_s^4}\right],
 \label{feq-disc-eq}
\end{equation}
where $\rho$ is the density and $\vec{u}$ is the macroscopic velocity of the fluid, calculated as follows:
\begin{equation}
 \rho = \sum_\alpha f_\alpha , \quad \vec{u} = \frac{1}{\rho} \sum _\alpha \vec{c}_\alpha f_\alpha  .
 \label{u-force-eq}
\end{equation}
The simulation results using LBM are expressed in lattice units, in which $\rho=1$, $\Delta t =1$ and $\Delta x=1$.

At the solid surfaces moving with velocity $\vec{u}_s$, the no-slip condition is applied though the bounce-back boundary condition:
\begin{equation}
 f_{\bar\alpha}(\vec{x}_s,t+\Delta t) = {f_{\alpha}}^\ast(\vec{x}_s,t) - 2w_\alpha \rho_s \frac{\vec{c}_\alpha \cdot \vec{u}_s}{c_s^2},
 \label{bounce-back-u-eq}
\end{equation}	
where $\bar \alpha = -\alpha$ represents the opposite velocity vector, ${f_{\alpha}}^\ast$ is the distribution propagated towards the solid surface and $\rho_s$ is the average density of the fluid node neighbors. The drag force on the solid is calculated using the momentum exchange method~\cite{ladd_1994, PhysRevE.65.041203}:
\begin{align}
 \vec{F_d} = \frac{\Delta x ^3}{\Delta t} \sum_{\vec{x}_s, \alpha} \left[   2{f_{\alpha}}^\ast(\vec{x}_s,t) - 2w_\alpha \rho_s \frac{\vec{c}_\alpha \cdot \vec{u}_s}{c_s^2} \right] \vec{c}_\alpha,
 \label{drag-force-eq}
\end{align}
which uses the reflected distributions in the bounce-back $f_{\alpha}^\ast$. When two particles approach each other and there are few or no fluid nodes between their surfaces, Eq.~\ref{drag-force-eq} becomes inaccurate since the space between the particles is interpreted as vacuum. To correct for this, we consider a hydrodynamic lubrication force, which is repulsive and points in the direction connecting the center of the particles of strength~\cite{IGLBERGER20081461}:
\begin{eqnarray}
 F_l = \frac{3\pi \rho \nu R^2 v_r}{2h},
\end{eqnarray}
where $v_r$ is the relative velocity between the particles, $h$ is the minimum distance between their surfaces and $R$ is the radius of the particles. This force diverges as $h\rightarrow 0$; thus we introduce a cutoff at $h=0.01R$. All the forces acting on the particles are combined to describe its trajectory using the Verlet method. As the particle moves in the fluid, liquid nodes are destroyed and created. For the destroyed fluid nodes, the fluid information is erased and the fluid momentum is transferred to the solid. For the new fluid nodes as in Ref.~\cite{aidun_lu_ding_1998}, the density is set to the average density of the neighbouring fluid nodes in the first belt $\bar \rho$ (considering all nodes reached by the lattice vectors of D3Q19) and the velocity field to the solid velocity $\vec{u}_s$. The distribution function of these new nodes is the corresponding equilibrium distribution, $f_\alpha = f^{eq}_\alpha(\bar \rho, \vec{u}_s)$.

The collisions between particles $i$ and $j$ are mediated by an elastic force~\cite{C6SM00357E}: 
\begin{eqnarray}
 \vec{F}^{el}_{ij} =  
\begin{cases}
    0,& \boldsymbol{\varepsilon}_{ij}=0\\
    k_l\boldsymbol{\varepsilon}_{ij},&  \frac{d\boldsymbol{\varepsilon}_{ij}}{dt} \geq 0  \\
    k_u\boldsymbol{\varepsilon}_{ij},& \frac{d\boldsymbol{\varepsilon}_{ij}}{dt} < 0  ,
\end{cases} 
\label{el-force-eq}
\end{eqnarray}
where $\boldsymbol{\varepsilon}_{ij}$ is the overlapping vector, defined as:
\begin{eqnarray}
 \boldsymbol{\varepsilon}_{ij} =  
\begin{cases}
    0,& \text{if } 2R - \vert \vec{r}_{ij}\vert  \leq 0\\
    2R - \vert \vec{r}_{ij}\vert  & \text{otherwise}.
\end{cases} 
\end{eqnarray}
Here, $\vec{r}_{ij}$ is the vector connecting the centers of the two particles
from $j$ to $i$ and $k_l$ and $k_u$ are the elastic coefficients when the
colliding particles are approaching or moving away from each other, respectively. The interaction with the walls also follows Eq.~\ref{el-force-eq}, but considering the distance of the particle to the wall.

In the LBM simulations, we use the following parameters, all in lattice units. The dimensions of the simulation box are $L_x \times L_y = 113 \times 413$ and the relaxation time is $\tau_f=0.7$. We simulate 100 circular particles with radius $R=5$, the same density as the fluid, $k_u=0.1$ and $k_s=0.2$, up to $t=4\times 10^6$, which is sufficient for all the particles to sediment and stop moving.  In order to consider particles with different terminal velocities, we apply different external forces to each species (fast and slow). To the slow particles we apply an acceleration (or force density) $a_s = 3\times 10^{-6}$ and, to the fast ones, $a_f = a_s/\beta$, where $\beta \leq 1$ is the ratio of the forces. As will be discussed in Sec.~\ref{lbm-result-sec} we find that $\beta$ corresponds to the ratio between the velocities of the slow $v_s$ and fast particles $v_f$ for single particles: $\beta = v_s/v_f$ . The particles of each species are chosen uniformly at random and we consider 10 samples for each value of $\beta$.

	\section{Results~\label{sec::results}}
	\begin{figure}
		\centering
		\includegraphics[width=\columnwidth]{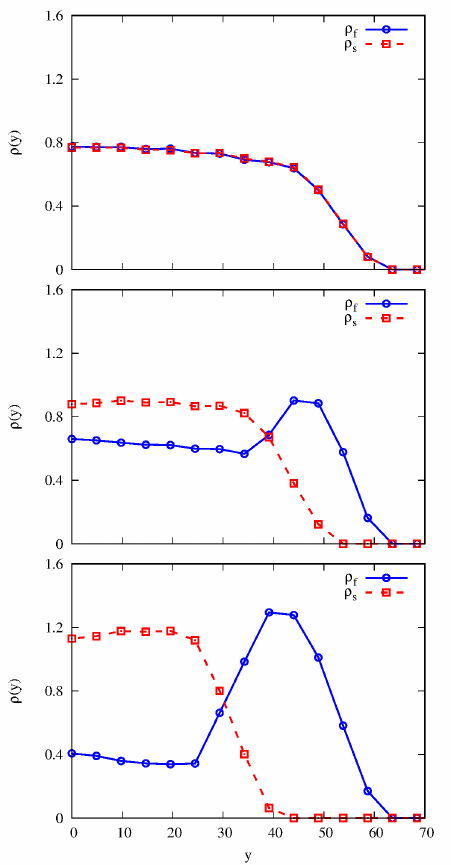}
		\caption{Density profiles as function of the height in the final deposit for fast (red lines) and slow (blue lines) particles averaged over $10^2$ samples for (a) $v=\frac{v_s}{v_f}=1$, (b) $v=0.5$ and (c) $v=0.1$, and $g=12$.}\label{fgr:densityprofiles}
	\end{figure}

	In the overdamped regime and neglecting thermal fluctuations, one expects that single particles move with a constant sedimentation velocity given by $v_{i}=\frac{mg}{\gamma_i}$. The rate of particle accumulation at the bottom depends on the flux density of each species at the growing front of the deposit, $\mathcal{J}_i=\rho_iv_i$, where $\rho_i$ is the particle number density of species $i$. Differences in the flux density of each species result in demixing along the vertical direction as particles accumulate at different rates on the bottom and thermal fluctuations are not strong enough to promote mixing. In what follows, we set the particle densities to be the same (equimolar mixture) and vary their velocities only. The demixing that occurs during sedimentation is, therefore, purely dynamical in nature. The relevant control parameter is the ratio between the sedimentation velocities of the two species, $v=\frac{v_s}{v_f}=\frac{\gamma_f}{\gamma_s}$. In order to investigate the dependence on this parameter, in the results that follow, we fix $\gamma_f$ and vary $v$ by changing $\gamma_s$, i.e., by changing $v_s$, with $v_f$ constant. The degree of demixing depends on $v$ as seen from the final deposit density profiles $\rho_f(y)$ and $\rho_s(y)$ in Fig.~\ref{fgr:densityprofiles}. When  $v=1$ the density profiles are identical as the particles are indistinguishable. When $v \neq 1$, the final deposit can be divided into two regions: one, at the bottom, where the density of the fast particles is higher than the density of the slow ones and another, at the top of the deposit, composed essentially by slow particles. The difference in the particle densities in the first region and the thickness of the second region increase as $v$ decreases (see Figs.~\ref{fgr:densityprofiles} $(b)$ and $(c)$). 

	\begin{figure}
		\centering
		\includegraphics[width=\columnwidth]{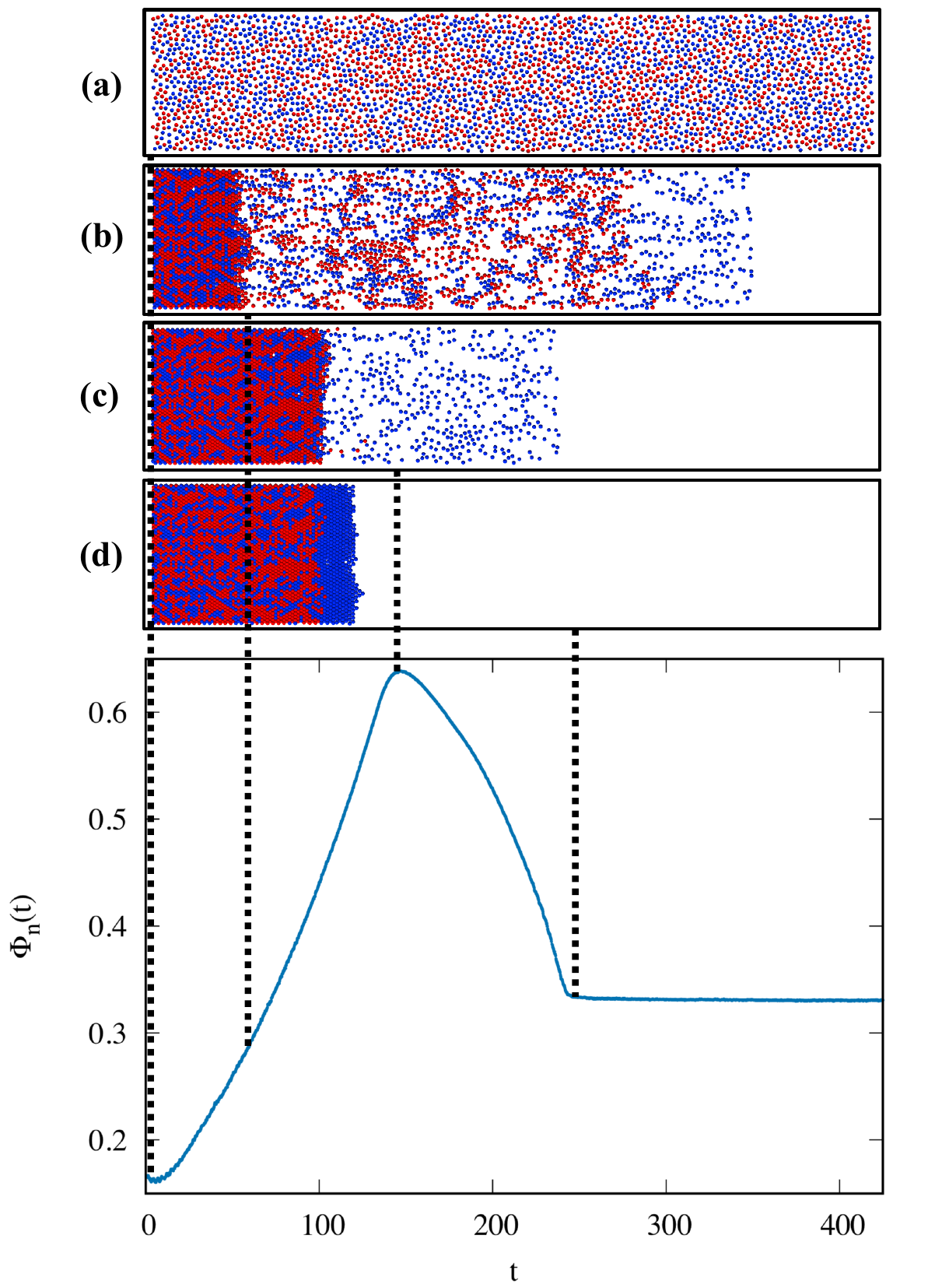}
		\caption{Time evolution of the parameter $\Phi_n$ averaged over $10^2$ samples for $v=\frac{v_s}{v_f}=0.5$, $g=12$ and $\rho_0=0.4$. The snapshots are for four different values of $t$, namely, (a) $0$,(b) $70$,(c) $150$, (d) $275$.}\label{fgr:snapshotphi}
	\end{figure}

	In order to characterize the segregation along the vertical direction, we define a parameter 
	\begin{equation}
	\Phi  = \frac{1}{L_{y}'}\int\limits_0^{L_{y}'} \frac{\lvert \rho_{f}-\rho_{s} \rvert}{\rho_{f}+\rho_{s}} dy,
	\label{phi}
	\end{equation} 
	where $L_y'$ corresponds to the height at which the last moving (slow) particle is located, given by $L_y'= L_y-v_st$. 

	\begin{figure}[ht]
		\centering
		\includegraphics[width=\columnwidth]{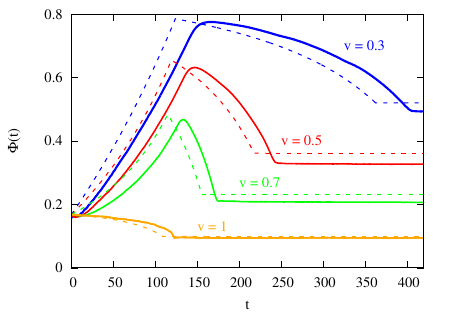}
		\caption{Time dependence of $\Phi_n(t)$ on the ratio of the particle velocities, $v$. The solid lines with open symbols are the results of BD simulations and the dashed lines are obtained from the analytical model (Eq. \ref{Eq:mean-field}).}\label{fgr:phiv}
	\end{figure}
	
	\subsection{Numerical results~\label{subsec::numericalresults}}
	
	To evaluate numerically the integral in Eq. (3), we divided the simulation domain into horizontal slices of height $\Delta y=1.5$ and width $L_x$. The integral is then converted into a sum,
	\begin{equation}
	\Phi_n(t)= \frac{1}{N_b}\sum_{i}^{N_b} \frac{\lvert N_{f} - N_{s} \rvert}{N_f + N_s},
	\label{phin}
	\end{equation}
	where $N_b$ is the number of slices and, $N_f$ and $N_s$ are the number of fast and slow particles in each slice, respectively. This parameter is one if the species are completely segregated and zero if they are perfectly mixed. 
	
	The time evolution of $\Phi_n$ is shown in Fig. \ref{fgr:snapshotphi} (bottom plot). Initially, $\Phi_n$ grows until it reaches a maximum, $\Phi_{max}$, at a time defined as $t^*$. For $t>t^*$, this parameter decreases until it saturates asymptotically. Note that $\Phi_n$ is not zero at $t=0$. Since the particles are initially distributed uniformly at random, the average absolute difference $\lvert N_{f}-N_{s} \rvert$ for a given slice can be estimated from a binomial distribution. This difference follows a half-normal distribution with mean $\mu=\sqrt{\frac{2N_t}{\pi}}$, where $N_t=N_f+N_s$ is the total number of particles in the slice. Therefore, the value of $\Phi_n$ at the starting configuration is $\Phi_n(0)=\Phi_0=\sqrt{\frac{2}{\rho_0L_x \Delta y \pi}}$, where $\rho_0$ is the initial total density of particles and it vanishes only in the thermodynamic limit. The initial increase in $\Phi_n$ corresponds to the dynamical demixing regime where there is a rapid accumulation of the fast particles in the deposit with a fraction of the slow particles dragged along while the remaining slow particles lag behind. The peak is reached when all the fast particles deposit at $t=t^*$ (see Fig. \ref{fgr:snapshotphi}$(c)$). For $t>t^*$, $\Phi_n$ decreases until the remaining slow particles deposit, on a top layer consisting (almost) exclusively of slow particles, if the difference in the velocities is sufficiently large (see Fig. \ref{fgr:snapshotphi}$(d)$). We define this instant as the saturation time, $t_{final}$. Note that, the level of segregation of the particles remains almost the same from $t^*$ to $t_{final}$, as seen from the snapshots $(c)$ and $(d)$. This parameter depends on the level of segregation and on $L'_y(t)$ that decreases with time. The region where only slow particles are present occupies a larger area at $t^*$ than at $t_{final}$ and the same number of particles contribute to the integral. 
	
	To characterize the structure of the deposit we measured the six-fold bond order parameter, $\langle \phi_6 \rangle$, defined  as
	
	\begin{equation}
	\langle \phi_6 \rangle = \frac{1}{N} \sum_{l}^{N} \frac{1}{6}\left| \sum_{j}^{N_b}e^{i6\theta_{lj}}\right| , \label{Eq:bond}
	\end{equation}
	where $N$ is the total number of particles, $N_b$ is the number of neighbors (within a cutoff of $2.5$) and $\theta_{lj}$ is the angle between the line that connects the particles $j$ and $l$ with the $x$-axis. This is a continuous order parameter that is one when the particles are arranged in a perfect hexagonal structure and it decreases when the order decreases. At the low temperatures considered, the particles in the deposit form a nearly perfect hexagonal structure with $\langle \phi_6 \rangle = 0.9$ and the level of segregation remains the same until the end of the simulation ($\Phi_n$ is constant). This does not correspond to the configuration expected at thermodynamic equilibrium where the particles form a completely mixed phase, as they are indistinguishable \cite{Nunes2018}. Obviously, the deposits observed at the end of the simulations are transient but their relaxation towards equilibrium occurs on much longer timescales than the observation time.
	
	The solid lines with open symbols in Fig. \ref{fgr:phiv} show the time dependence of $\Phi$ obtained numerically for different values of the velocity ratio $v$. As expected, $\Phi_{max}$ increases as the ratio $v$ decreases from one, revealing that, as the difference of the sedimentation velocities increases, higher levels of segregation are attained in the deposit. 
	
	Reducing $v$, decreases the sedimentation velocity of the slow particles (keeping the velocity of the fast particles constant). However, the effective velocity of the fast particles also decreases due to the interactions with the slow ones (which act as obstacles) and, as a consequence, the time when the peak in $\Phi$ is reached, $t^*$, also increases. The saturation time, $t_{final}$, is also affected by $v$ since the difference between $t^*$ and $t_{final}$ is the time taken by the remaining slow particles to deposit and this depends only on their sedimentation velocity. 
	
\begin{figure}
		\centering
		\includegraphics[width=\columnwidth]{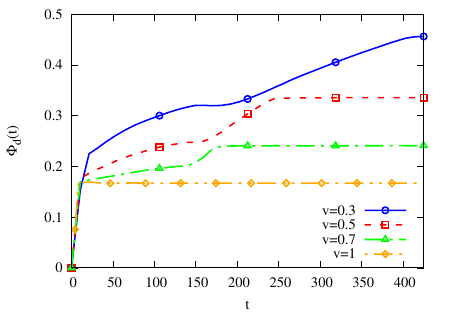}
		\caption{Time evolution of the segregation in the deposit.}\label{fgr:phi_deposit}
	\end{figure}

	The parameter $\Phi_n$ measures the segregation in the entire system but we can also measure the segregation in the deposit using the following definition~\cite{Rex2007}
	\begin{equation}
	\Phi_d= \frac{1}{N_d}\sum_{i}^{N_d} \frac{\left( N_{f} - N_{s} \right)^{2}}{\left(N_f + N_s\right)^{2}},
	\label{phi_gl}
	\end{equation}
	where $N_d$ is the total number of particles with $\phi_6>0.9$. $N_f$ and $N_s$ are the number of fast and slow particles within the cutoff distance $2.5$ of particle $i$. The segregation in the deposit increases monotonically with time (see Fig.~\ref{fgr:phi_deposit}). At $t^*$, the slope of the curve changes signalling the phase where only slow particles arrive.

	\subsection{Analytical model~\label{subsec::mean-field}}
	\begin{figure*}[ht]
		\centering
		\includegraphics[width=15cm]{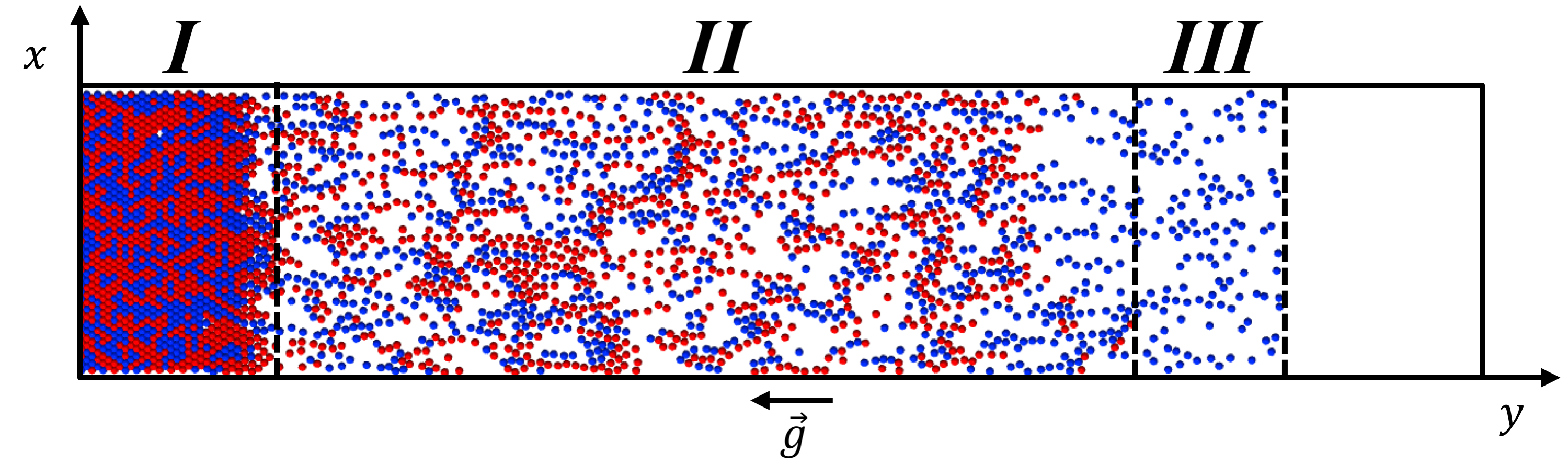}
		\caption{To evaluate the integral in Eq. \ref{phi} the space is divided into three regions: region \textit{I} in the interval $[0,l^*(t)]$, region \textit{II} in $ [l^*,L_y-v_ft] $ and \textit{III} in $[L_y-v_ft, L_y-v_st]$. In region \textit{I} the integrand is given by Eq. \ref{density}, in region \textit{II} the integrand is $\Phi_0$ and we consider that the number of fast and slow particles entering this region is approximately the same as the number of particles leaving the region, and in region \textit{III} the integrand is one, as there are only slow particles. }\label{fgr:snapshot_integral}
	\end{figure*}
	
	We consider now a simple analytical model. We assume that the particles move with a constant sedimentation velocity $v_{i} = \frac{m g}{\gamma_i}$ that depends only on the particle species, and we neglect particle-particle interactions and thermal fluctuations. We define the thickness of the packed deposit as $l^*(t)$. The number of particles of a species in the region $y<l^*(t)$ at a given time is the number of particles initially at $y<l^*(t)$ plus those of that species that entered into that region. The latter can be estimated considering that the fraction of particles of species $i$ that entered into the region is  $\frac{v_i}{v_f+v_s}$. Despite the fact that particle interactions are soft, we assume an upper bound for the density, given by the packing fraction of disks with diameter $\sigma=1$, i.e.,  $\rho_{max}= \frac{2\sqrt{3}}{3}$. We can then estimate the density of particles by
	\begin{equation}
	\rho_i(y<l^*,t)= \frac{\rho_0}{2} + (\rho_{max}-\rho_0)\frac{v_i}{v_f+v_s},
	\label{density}
	\end{equation}
	where $(\rho_{max}-\rho_0)$ is an estimate of the increase in the local density.
	
	The integral in Eq. \ref{phi} for $t<t^*$ can be replaced by the sum of three terms, corresponding to the contribution of three different regions, as shown in Fig. \ref{fgr:snapshot_integral}. Region \textit{I} corresponds to the deposit, where the density of each species is given by Eq. \ref{density}. We also account for the non-zero segregation factor, $\Phi_0$, which arises from the initial uniform distribution of the particles and the discrete nature of the numerical integration. Region \textit{II}, where the two types of particles are perfectly mixed, is delimited by the surface of the deposit, $y \approx l^*(t)$ and  $y \approx L_y-v_ft$, the height of the last fast particle. Here, we assume that the number of particles of either species that leave region \textit{II} is approximately the same as the number that enters it and, therefore, the integrand is the constant $\Phi_0$. Region \textit{III} contains particles of only one type and is delimited by $y \approx L_y-v_st \equiv L'_y$, the position of the last slow particle. In this region, the integrand is one as only slow particles are present. The number of fast, $N_f$, and slow particles, $N_s$,  in regions \textit{I} and \textit{II} can then be regarded as the number of successes and failures in a binomial distribution of $N_t$ trials. The probability that a particle in each of these regions is a fast particle is
	\begin{equation}
	p_{f,\alpha}= \frac{\frac{\rho_0}{2}+(\rho_\alpha-\rho_0)\frac{1}{1+v}}{\rho_\alpha},
	\label{pf}
	\end{equation}
	and the probability that it is a slow particle is
	\begin{equation}
	p_{s,\alpha}= \frac{\frac{\rho_0}{2}+(\rho_\alpha-\rho_0)\frac{v}{1+v}}{\rho_\alpha},
	\label{ps}
	\end{equation}
	where $\rho_\alpha$ is the the total density of each region ($\rho_{max}$ in region \textit{I} and $\rho_0$ in region \textit{II}). The term $\lvert  N_{f} - N_{s}\rvert$ corresponds to the absolute difference between successes and failures of $N_t$ trials that follow a folded normal distribution whose expected value is given by: 
	  \begin{equation}
	\mu'_\alpha = \sigma_\alpha \sqrt{\frac{2}{\pi}} \exp\left(\frac{-\mu_\alpha^2}{2 \sigma_\alpha^2}\right) + \mu_\alpha \erf\left(\frac{\mu_\alpha}{\sqrt{2 \sigma_\alpha}}\right).
	 \end{equation}
	Here,  $\mu_\alpha = N_t(p_{f,\alpha}-p_{s,\alpha})$ and $\sigma_\alpha^2 =4 N_tp_{f,\alpha} p_{s,\alpha} + \mu_\alpha$ correspond to the expected value and to the variance of the Gaussian distribution (not ``folded'') of the difference between successes and failures for the probabilities $p_{f,\alpha}$ and $p_{s,\alpha}$. 
	The integral for $t \leq t^*$ is then approximated by
	\begin{equation}
	\begin{split}
	\Phi(t \leq t^*) = \frac{1}{L_{y}'} &\left[ \int\limits_0^{l^*(t)}\frac{\mu'_{I}}{\rho_{max} \Delta_y L_x} \ dy +\right.\\
	&\left. + \int\limits_{l^{*}}^{L_{y}-v_ft} \frac{\mu'_{II}}{\rho_{0} \Delta_y L_x} dy  +  \int\limits_{L_y-v_ft}^{L_{y}'} dy \right].
	\end{split}  
	\label{integral1}
	\end{equation}
	For $t^*<t<t_{final}$ the thickness of region \textit{II} is zero and then
	\begin{equation}
	\begin{split}
	\Phi(t > t^*) = \frac{1}{L_{y}'} &\left[  \int\limits_0^{L^*} \frac{\mu'_{I}}{\rho_{max} \Delta_y L_x} dy  +\int\limits_{L^*}^{L_{y}'} dy \right], 
	\end{split}  
	\label{integral2}
	\end{equation}
	where $L^*=l^*(t^*)$ is the length of the structure at $t=t^*$.
	
	We estimate $l^*(t)$ in the following way. The number of deposited particles is given by the flux of particles through the line defined by $y=l^*(t)$ plus the number of particles that is initially present in the region below this height. We consider that particles travel with constant velocity, $v_i$, for $y>l^*(t)$ and that the flux through that line for each species can be written as $j_i(t)=\frac{\rho_0}{2}v_iL_xt$. The number of particles in the deposit, $N_s$, is
	\begin{equation}
	N_s=\rho_{max} L_x l^*(t) = L_x l^*(t)  \rho_0+ \left(\frac{\rho_0}{2} v_f+\frac{\rho_0}{2} v_s \right) t L_x.
	\label{fast}
	\end{equation}
	We can now rearrange the terms and the expression for $l^*$ is
	\begin{equation}
	l^*(t) = \frac{\frac{\rho_0}{2}(1+ v)}{\rho_{max}-\rho_0}v_ft.
	\label{l}
	\end{equation}
	For $t>t^*$,  all the fast particles are deposited and
	\begin{equation}
	L^* = \frac{ \rho_0L_y (1+v)}{2\Big(\rho_{max}-\rho_0+ \frac{\rho_0}{2}(1+v)\Big)}.
	\label{L}
	\end{equation}
	Since $L^*=l^*(t^*)$,
	\begin{equation}
	t^* = \frac{(\rho_{max}-\rho_0)L_y}{v_f\Big[\frac{1}{2} \rho_0(1+v)+(\rho_{max}-\rho_0)   \Big]}.
	\label{t}
	\end{equation}
	The height of the final deposit is $l^*(t_{final})=\frac{N}{\rho_{max}L_x}$, where $t_{final}$ is the time when the last slow particle deposits, which is estimated as $t_{final}=(L_y-\frac{N}{\rho_{max}L_x})/v_s $. At later times the parameter $\Phi$ remains constant as no more particles are added to the deposit.
	
	Thus, the parameter $\Phi$ is
	\small
	\begin{equation}
	\Phi(t)=
	\begin{cases} 
	\frac{1}{L_y-v_st} \Big[ \frac{\rho_0(1+v)}{2 (\rho_{max}-\rho_0)}\Big(\frac{\mu'_{I}\rho_0-\mu'_{II} \rho_{max}}{\rho_{max} \rho_0 \Delta y L_x}-\\ \hspace{0.7cm} -\frac{\mu'_{II})}{\rho_0 \Delta y L_x}+1 - v\Big) v_f t +
	  \frac{L_y \mu'_{II}}{\rho_0 \Delta t L_x}  \Big], \hspace{.4cm} t \leq t^* \\ 
	\\
	
	1+\frac{L^*}{L_y-v_st}\Big[\frac{\mu'_{I}}{\rho_{max}\Delta y L_x} -1\Big], \hspace{0.3cm} t^* \leq t \leq t_{final}  \\
	\\
	1+\frac{L^*}{L_y-v_st_{final}}\Big[\frac{\mu'_{I}}{\rho_{max}\Delta y L_x} -1\Big],   \hspace{0.3cm} t \geq t_{final} .
	\end{cases}
	\label{Eq:mean-field}
	\end{equation}
	\normalsize

	The dashed lines in Fig.~\ref{fgr:phiv} show the time dependence of $\Phi$ obtained from Eq.~\ref{Eq:mean-field} for different values of the ratio $v$. The comparison with the numerical results reveals that the analytical calculation reproduces the main features of the simulations. The highest levels of segregation are achieved for the lowest $v$, when the difference between the velocities is largest. In this limit, the slow particles travel much slower than the fast ones and are, on average, the last to deposit forming a thick layer on top of the first deposit consisting of slow particles only. For $v=1$, the particles are indistinguishable and the maximum value of $\Phi$ is the (finite-size) initial value. 

	\begin{figure}
		\centering
		\includegraphics[width=\columnwidth]{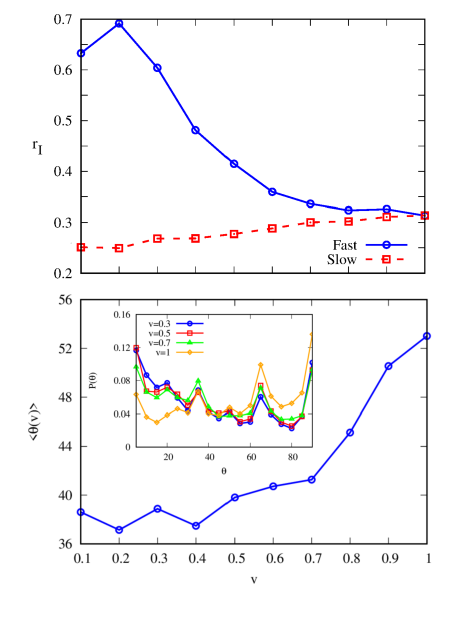}
		\caption{Top panel: ratio between the lowest and highest (non-zero) eigenvalues of the inertia tensor of the clusters of particles in the final deposit. Bottom panel: average angle between the principal eigenvector of the slow cluster with the $y$-direction. In the inset we show the cluster angle distribution for different velocity ratios.}\label{fgr:inertiamoment}
	\end{figure}

	So far, we discussed segregation along the $y$-direction. However, the snapshot in Fig.~\ref{fgr:snapshotphi}$(d)$ suggests that at the bottom of the deposit there are linear-like clusters along the $y$-direction, which may promote segregation along the $x$-direction. To further investigate this question, we identified all the clusters of particles in the final deposit below the top layer of slow particles and calculated their inertia tensor. We defined the parameter $r_I = \frac{1}{N_c}\sum r_c$, where $r_c$ is the ratio of the smallest and largest (non-zero) eigenvalues of the inertia tensor of cluster $c$ and the sum is over all clusters of the same species with $N_c$ the number of such clusters. This parameter is one when all the clusters are symmetric and falls below one as the clusters become asymmetric. We measure $r_I$ for clusters larger than two particles excluding the largest cluster, where finite size effects may be significant. A deposit with randomly distributed particles is obtained for $v=1$, when the particle are indistinguishable. In this limit, $r_I$ is low, which suggests the prevalence of symmetric clusters. In the limit of low $v$, the number of fast particles is much larger than the number of slow ones in the region under consideration and they form a single large cluster. Accordingly, as shown in Fig.~\ref{fgr:inertiamoment} (top panel), $r_I$ increases for fast particles as $v$ decreases. The opposite occurs for the slow particles where $r_I$ decreases, showing that, as the ratio of velocities decreases, the clusters of slow particles become asymmetric. In the bottom panel of Fig. \ref{fgr:inertiamoment}, we plot the average angle, $\langle\theta \rangle$, with the y-direction of the eigenvector corresponding to the largest eigenvalue of the inertia tensor for each cluster of slow particles. As $v$ decreases, $\langle\theta \rangle$ also decreases revealing that the clusters in the deposit tend to extend along the $y$-direction as the difference of the particles velocities increases. Accordingly, the distribution of the angles of the clusters with the $y$-direction (inset of Fig. 6) the number of clusters with small angles increases for low $v$. This is likely a consequence of the laning phenomenon observed in binary mixtures of species moving at different velocities in region \textit{II}~\cite{Rex2007}.  
	
		\begin{figure}
		\centering
		\includegraphics[width=\columnwidth]{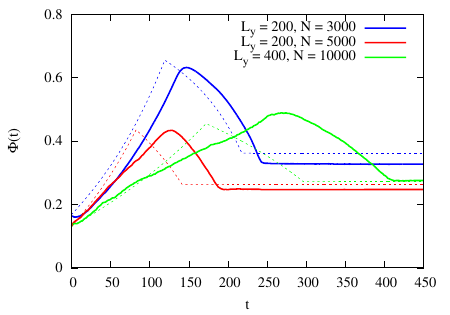}
		\caption{Order parameter from the simulations (solid line) and the analytical model (dashed line). The blue lines correspond to $v=0.5$ shown in Fig.~\ref{fgr:phiv}. The red and green lines illustrate the evolution of system with $60\%$ higher density. The red and blue lines are for systems with the same size while the green lines are for a system twice as large.}\label{fgr:highdensities}
	\end{figure}
	
	Although we have neglected the effect of particle-particle collisions during sedimentation in the analytical model, the simulation and analytic results are in good agreement for intermediate and low densities. In the limit of high densities, however, the simulation and analytic results disagree, as particle-particle interactions becomes relevant. In Fig.~\ref{fgr:highdensities} we plot the time-dependence of the order parameter for $v=0.5$ for the system depicted in Fig.~\ref{fgr:phiv} (in blue) with density $\rho_0 = 0.4$ and for a system with a density $60\%$ higher, for two system sizes (red and green). In spite of the fact that at higher densities the time dependence of the order parameter obtained from Eq.~\ref{Eq:mean-field} is not in quantitative agreement with the numerical simulations, the maximum of the order parameter and its final value are similar.

	\subsection{Hydrodynamic effects}
	\label{lbm-result-sec}
	
We consider now the role of hydrodynamics in the sedimentation of particles with different velocities. As described in Sec.~\ref{lbm-intro-sec}, we simulate the fluid flow and fluid-particle interaction using the LBM and the particle sedimentation is driven by an external force $\vec{a}$, with different values for the two species (fast and slow). 

    \begin{figure}
		\centering
		\includegraphics[width=0.8\columnwidth]{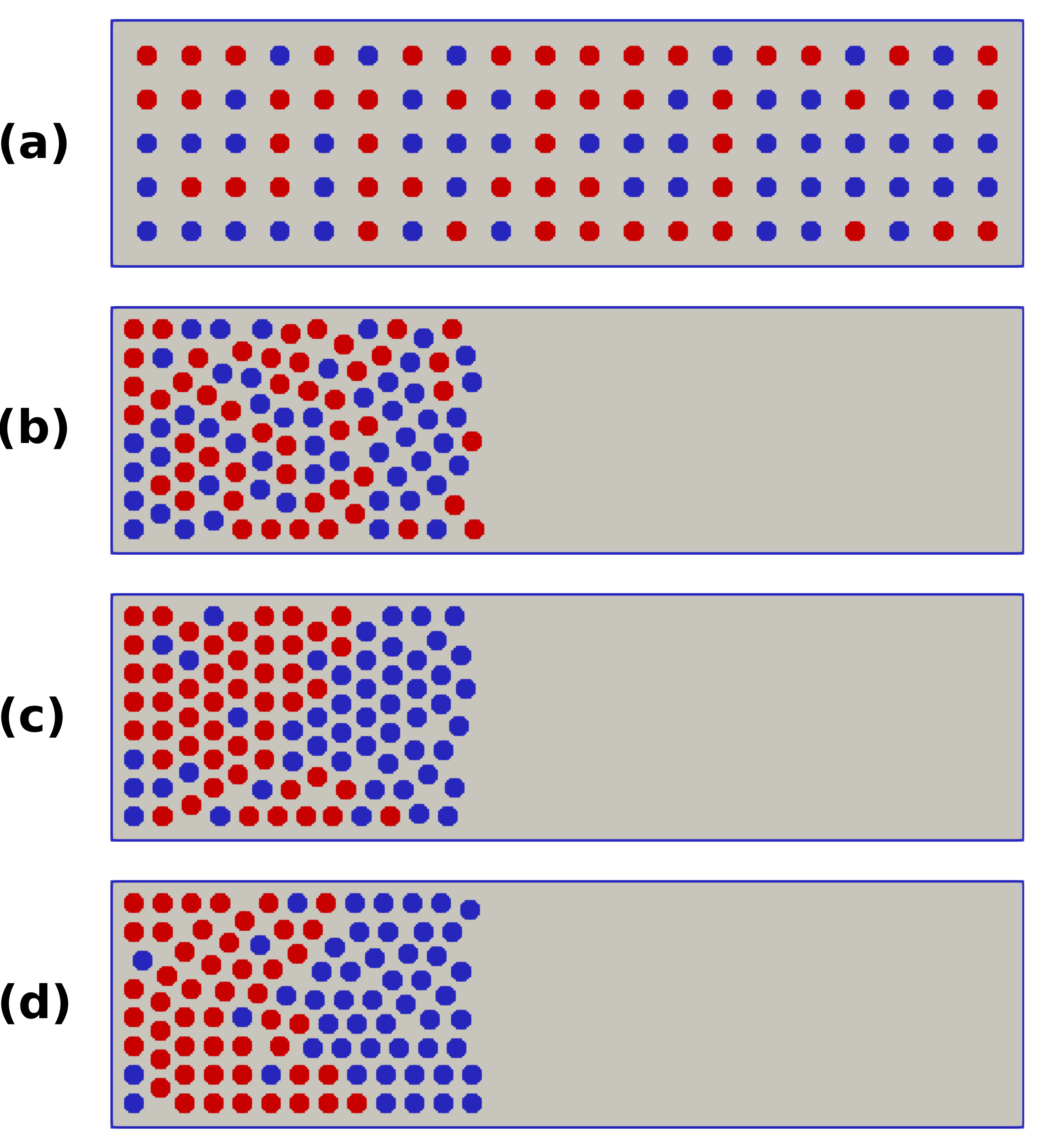}
		\caption{Results from the LBM. (a) Initial configuration for one of the samples. Final configuration for (b) $\beta=1$, (c) $\beta=0.5$ and (d) $\beta=0.1$. The fast particles are red and the slow ones blue.}\label{fgr:lbm1}
	\end{figure}

    \begin{figure}
		\centering
		\includegraphics[width=0.8\columnwidth]{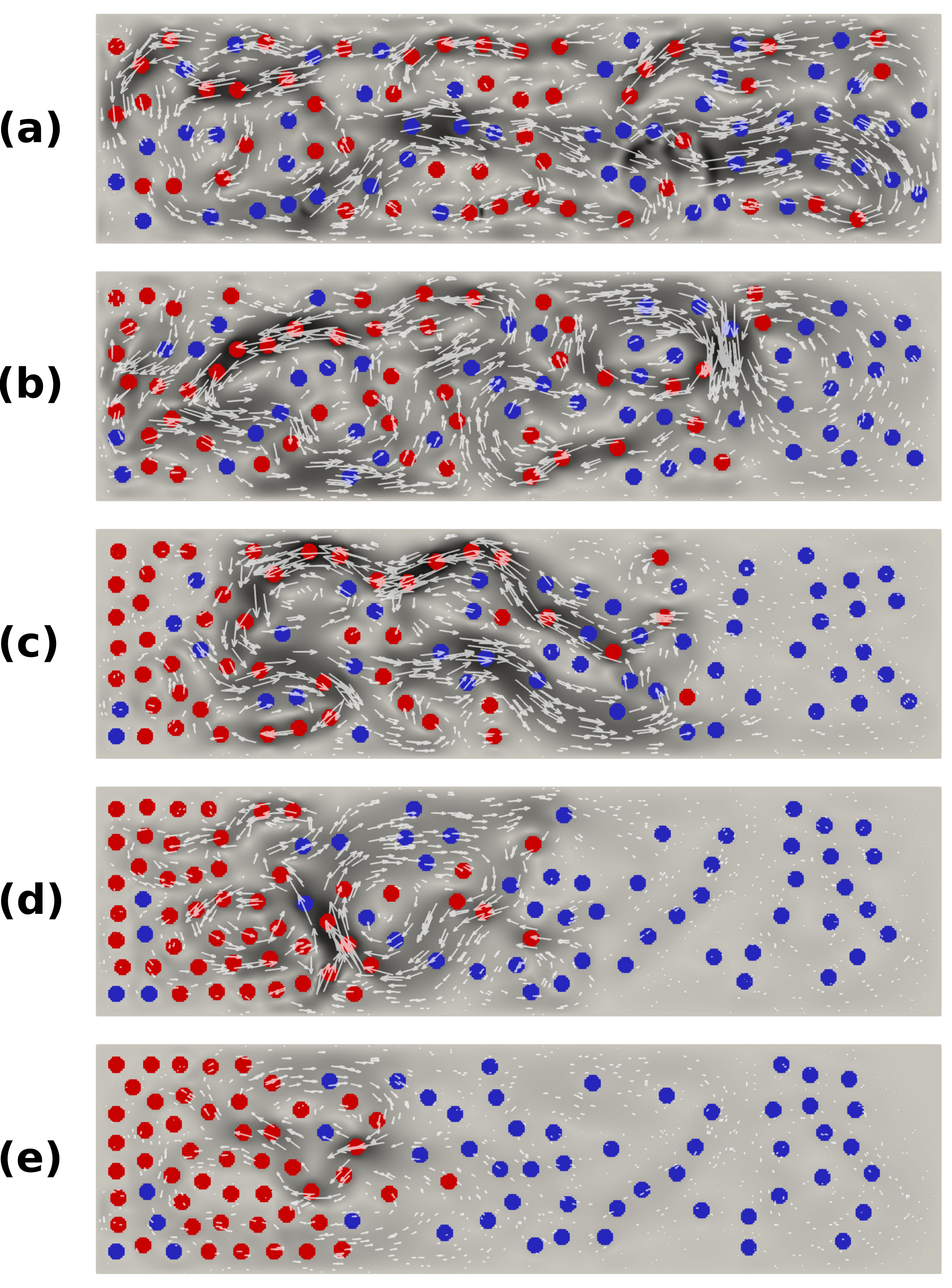}
		\caption{Velocity field (using the LBM) for the sample shown in Fig.~\ref{fgr:lbm1}(a) with $\beta=0.1$ at different times: (a) $t=10000$, (b) $t=40000$, (c) $t=70000$, (d) $t=120000$ and (e) $t=160000$. The fast particles are red and the slow ones blue. The background in grayscale represents the magnitude of the fluid velocity, with light gray corresponding to $u=0$ and black to $u=0.0048$. The arrows indicate the direction of the velocity vector.}\label{fgr:lbm2}
	\end{figure}
	
Figure~\ref{fgr:lbm1} shows the initial configuration of one of the samples and its final state for three different force ratios $\beta=\frac{a_s}{a_f}$. From visual inspection, we notice that for $\beta=0.1$, where the fast particles have the highest velocity, particles in the final deposit are more segregated than for the other two values of $\beta$. This behaviour is in line with what we found without hydrodynamics. When the time evolution of the particles is analyzed, as shown in Fig.~\ref{fgr:lbm2}, some major differences are revealed, which result from hydrodynamic effects. For example, we observed Rayleigh-Taylor instabilities \cite{Wysocki2009, Milinkovic2011, Huh2007}  as shown in Fig.~\ref{fgr:lbm2}. Initially, in order for the particles to go down, the fluid below them has to go up generating upcurrents. Despite the sedimentation force acting on the particles, these upcurrents can move the particles up, by contrast to the results of the MD simulations and the analytical model. Thus, quantitative differences due to hydrodynamics are to be expected.

	\begin{figure}
		\centering
		\includegraphics[width=0.9\columnwidth]{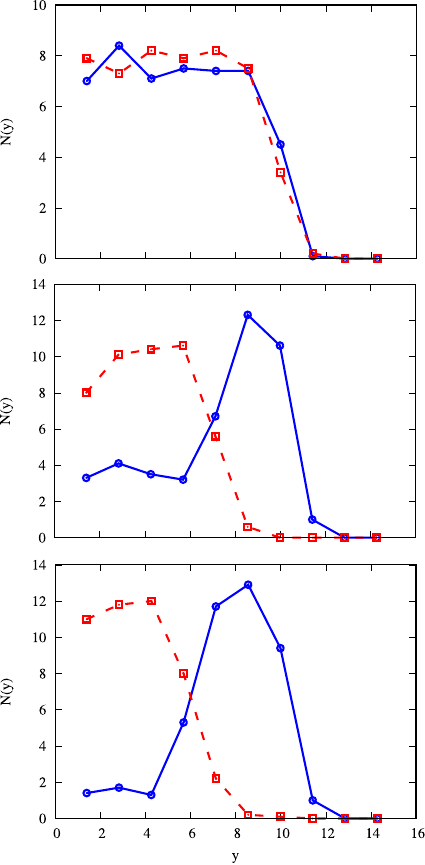}
		\caption{Number of particles in a given bin as function of the height in the final deposit for fast (red squares) and slow (blue circles) particles averaged over $10$ samples for (a) $\beta=\frac{a_s}{a_f}=1$, (b) $\beta=0.5$ and (c) $\beta=0.1$. Results from the LBM. }\label{fgr:density}
	\end{figure}
	
	\begin{figure}
		\centering
		\includegraphics[width=\columnwidth]{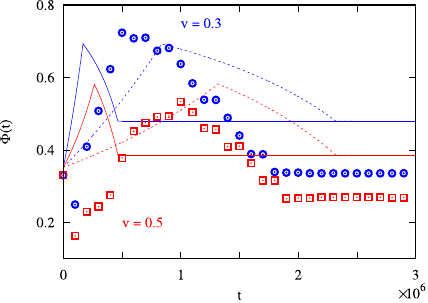}
		\caption{Time dependence of the parameter $\Phi_n(t)$ for two values of the ratio of the particle forces $\beta=\frac{a_s}{a_f}$. The open symbols are results of the LBM and the lines are from the analytical model (Eq.\ref{Eq:mean-field}). The solid lines use the sedimentation velocity of a single particle while the dashed line uses this velocity divided by five, which is approximately the average velocity of all particles in the sedimentation simulation with 100 particles.}\label{fgr:phi}
	\end{figure}

In Fig.~\ref{fgr:density}, we show the profile of the number of particles as a function of the height for fast and slow particles for different values of $\beta$. 
This corresponds to the final states, shown in Fig.~\ref{fgr:lbm1}, but now averaged over 10 independent samples. The behaviour of the profiles is qualitatively the same as those shown in Fig.~\ref{fgr:densityprofiles}. Segregation is observed for the system with $\beta=0.1$, and no segregation occurs for $\beta=1$ (equal sedimentation velocities for both species). In Fig.~\ref{fgr:phi}, we compare the time evolution of the order parameter $\Phi$ (Eq.~~\eqref{Eq:mean-field}) from the LBM simulations with the analytical model. To obtain the sedimentation velocity using the analytical model, we simulated the sedimentation of a single particle for different forces and measured the terminal velocity. In this case, the velocity increases linearly with the applied force, as expected for the values of the Reynolds numbers considered. Thus, the ratio of the forces is equal to the ratio of the velocities: $\beta = v$. The analytical curves using this velocity are represented by the solid lines in Fig.~\ref{fgr:phi}. We find that the sedimentation process for this velocity is much faster without hydrodynamics. This difference is due the upwards fluid displacement when the particles are sedimenting, which reduces their velocity (and even inverts the direction of motion). Thus, the average velocity of all particles is much lower due to the retarding effects of hydrodynamics. Numerically, we observe that the average velocity of particles of one specie (fast or slow) is nearly one fifth the velocity of a single particle of the same specie sedimenting in the fluid. 

We also show in Fig.~\ref{fgr:phi} the analytical results using this average velocity (dashed lines). Changing the velocity changes 
the time scale of the sedimentation, which becomes closer to the time scale of the LBM simulations. The final value of the order parameter is different from that predicted analytically as the velocity is highly non-uniform, which is not considered in the analytical model. However, the existence of a maximum of the order parameter is captured by the analytical model.

	\section{Conclusions~\label{sec::conclusions}}
	We investigated the dynamics of sedimentation of a simple binary mixture, and observed purely dynamical demixing. 
	
	We started by considering that the two species differ in their Stokes coefficients only, i.e., they have different sedimentation velocities. Since the species travel at different velocities, they demix dynamically as they move towards the bottom of the container in a gravitational field. We measured the degree of demixing without hydrodynamics using Brownian dynamics simulations for different ratios of the velocities and proposed a simple analytical description in the low density limit. We found that the analytical model captures the dynamics and the degree of segregation in the system even though particle-particle interactions are not taken into account. We also considered the sedimenting of particles in a hydrodynamic setting using the lattice Boltzmann method and found similar qualitative behavior, in particular, the existence of purely dynamical demixing. Quantitative differences, however, were observed resulting from the currents set up during the sedimentation which strongly affect the particles velocities, making them non-uniform.
	
	We focused on equimolar binary mixtures but the same demixing mechanism will occur for mixtures with any other composition. In fact, the composition of the deposit is determined by the ratio of the fluxes of sedimenting of the two species and, as a result, the initial composition of the mixture will affect the composition of the final deposit. The mechanism described here could also be used to obtain fully mixed deposits, since by tuning $v$ an equimolar deposit may be formed from mixtures poor in one of the two components.
	
	As a final note, we focused on colloidal suspensions but our conclusions can be extended to systems with larger particles, as we considered the limit of high P\'{e}clet number, where the dominant mechanism of mass transport is advection and thermal fluctuations are negligible. Finally, since in this limit the relevant mechanisms depend only on the ratio of sedimentation velocities, we expect the same behavior for particles with the same shape but different buoyant masses. The nature of the field is also irrelevant, and thus similar results are to be expected for other (constant) external fields (e.g., electromagnetic field).
	
	\section{Acknowledgements}
	
	We acknowledge financial support from the Portuguese Foundation for Science and Technology
	(FCT) under the contracts no. EXCL/FIS-NAN/0083/2012, SFRH/BD/119240/2016, UIDB/00618/2020 and UIDP/00618/2020.

	\bibliography{refs}
\end{document}